\documentclass[fleqn,12pt,twoside]{article}
\usepackage{espcrc1}
\usepackage{graphicx}
\usepackage[figuresright]{rotating}

\newcommand{\AmS}{{\protect\the\textfont2
  A\kern-.1667em\lower.5ex\hbox{M}\kern-.125emS}}

\title{Experimental determination of the Townsend coefficient for Argon-CO$_{2}$ gas mixtures at high fields}

\author{G. Auriemma\address[USB]{Universit\`{a} degli Studi della Basilicata, Potenza, Italy
and I.N.F.N. Sezione di Roma I, Rome, Italy}, D.
Fidanza\addressmark[USB], G. Pirozzi\addressmark[USB] and C.
Satriano\addressmark[USB] }

\begin{document}

\maketitle

\begin{abstract}
The first Townsend coefficient for Ar-CO$_2$ based gas mixtures
has been measured over a wide range of reduced electric field. The
experimental setup and the measurement technique are described
here. A linear superposition model has also been successfully
applied. \vspace{1pc}

\bigskip
PACS-1996  29.40.Cs - 52.80.Dy


\end{abstract}

\section{Introduction}

The Townsend coefficient, defined as the number of ions produced per unit path
by a single electron traversing a gaseous medium, is of fundamental importance
in all the discharge
processes, in particular in the description of the electronic gain of gas
ionization detectors\cite{Sauli:2000em} or plasma discharge devices\cite{veronis2000}.
Direct measurements of the Townsend coefficient are poorly
reported in the literature\cite{Dodokhov:1980qv,Sharma:1992ge,Sharma:1993wp,2000JAPh...88...6192K,Arefev:1993mr},
therefore the most used estimate of the Townsend coefficient is based upon a numerical solution of the Boltzmann
equation\cite{Biagi:1989rm,2000JPhD...33...62U}.

In this paper we report the measurement of the effective Townsend
coefficient for Ar-CO$_2$ gas mixtures in the typical working
condition of high gain MWPC \cite{LHCbTDR:2001}. We have derived the effective
Townsend coefficient $\alpha$ from the measurements of the gas
gain $M$ in a cylindrical test tube, using a new method that is
briefly described in the following.

The gas gain $M$, defined as the ratio of the anode current $I$
over the primary ionization current $I_0$ that would be measured
operating the tube in the ionization chamber mode, can be
calculated integrating the Townsend coefficient from the starting
point of the avalanche $r_0$ to the surface of the
wire\cite{Rose:1941}. In practice we have
\begin{eqnarray}\label{eq:intro1}
\ln M & = & \int_{r_{0}}^{r_{a}}\,\alpha(S)\, dr
\end{eqnarray}
where $S$ is the reduced field, which in cylindrical geometry will
be\begin{equation}\label{eq:intro2}
    S(r)=\frac{V}{p\,r\,\ln (r_c/r_a)}
\end{equation}
being $V$ the voltage applied to the anode, $r_a$ the anode wire
radius, $r_c$ the cathode radius and $p$ the pressure of the gas.
It is worth noticing that the coefficient $\alpha$ which appears
in Eq.\ (\ref{eq:intro1}) is an effective coefficient in the sense
that it is the difference between the absolute coefficient and the
attachment coefficient. Changing the integration
variable\cite{2000JAPh...88...6192K} and using  Eq.\
(\ref{eq:intro2}), we can recast Eq.\ (\ref{eq:intro1}) in the
form
\begin{equation}  \label{eq:method3}
\ln M=\frac{V}{p\,\ln
(r_c/r_a)}\,\int_{S_{0}}^{S_{a}}\,\alpha(S)\,\frac{dS}{S^2}
\end{equation}
where $S_{a}=S(r_a)$ and $S_{0}=S(r_0)$. A small change in $V$ is
equivalent to a small change in the value of $S_a$.
Differentiating Eq.\@ (\ref{eq:method3}) respect to $S_a$, treated
as an independent variable,we obtain after some algebraic manipulations
\begin{equation}  \label{eq:alpha}
\alpha(S_a) = \frac{1}{r_a}\left(\frac{d \ln M}{d \ln V} - \ln M
\right)
\end{equation}
which we have used to extract from our measurements the Townsend
coefficient. It is noteworthy that, being always $\ln M\ll d \ln
M/d \ln V$, Eq.\@ (\ref{eq:alpha}) can be used to estimate
$\alpha$ even if the primary ionization current $I_0$ is poorly
known, because $d\ln M\equiv d\ln I$ if $I_0$ is constant.

\section{Experimental setup}

\begin{figure}[ht!]
\centering
\includegraphics[width=8cm]{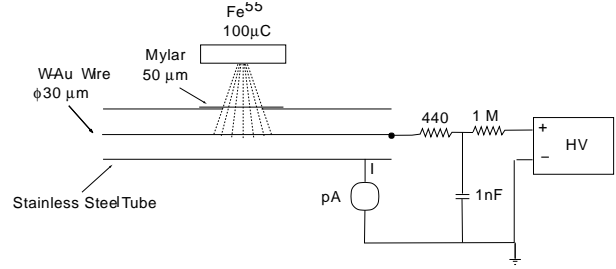}
\caption{Schematic view of the experimental set up}
\label{fig:tube}
\end{figure}

The experimental setup used in this work is shown in Fig.\@\ref{fig:tube}.
The active counting volume is a cylinder with a thin anode wire of gold
plated tungsten of 30 $\mu$m diameter, soft soldered to the bronze
feed-throughs and subjected to a mechanical tension of $60$ g. The wire is
accurately located in the center of a precision stainless steel tube of $%
(5.64 \pm 0.01)$ mm inner diameter and $(0.18 \pm 0.01)$ mm wall
thickness. Research grade pure gases were mixed, with relative
percentages controlled to $0.5 \%$ accuracy level by two computer
driven mass flowmeters; during operation the test tube has been
fluxed with a total flow of 10 sccm. The high voltage is given by
a power supply device, controlled with a IEEE-488 interface, while
the current is read with a Keithley Mod.\@ 485 picoammeter, using
the internal logarithmic conversion and remotely controlling the
fullscale. The calibration error of the picoammeter stated by the
constructor is 0.4\% plus 1/2 LSD of the readout.

The active volume of the tube has been exposed to a radioactive
$^{55}$Fe source, whose intensity was 100 $\mu$Ci. The X-rays
reach the gas volume by passing through a small mylar window of
50 $\mu$m thickness placed on the cylinder external surface. The
distance of the source from the mylar window has been adjusted to
give a rate of current pulses from X-rays interactions in the
active volume of the tube of about 100 kHz. The primary ionization
current $I_0$ has not been directly measured. Instead it has been
calculated from the above quoted rate multiplied by the average
primary charge released in the counter by the 5.9 keV X-rays,
taking into account the escape for argon. As we said in the
discussion of Eq\ (\ref{eq:alpha}) the uncertainty in the
determination of $I_0$ has practically no effect on the calculated
value of $\alpha$. However we observe that also if the flux of
X-rays from the radioactive source is constant, the counting rate
will depend upon the relative composition of the gas mixture, due
to the different absorption of X-rays in the two gases, and the
absolute pressure and temperature which determine the density of
the gas. In fact the primary ionization current $I_0$ will be
\begin{eqnarray}\label{eq:current}
I_0 & \propto & \left(1-e^{-\mu_{Ar}\,x_{Ar}}\right) \cdot  \nonumber \\
& \cdot & \frac{E_X (1-P_{esc})+(E_X - E_{K}^{Ar})P_{esc}}{w_{Ar}} +
\nonumber \\
& + & \left(1-e^{-\mu_{CO_2}\,x_{CO_2}}\right)\frac{E_X
}{w_{CO_2}}
\end{eqnarray}
where $\mu$ are the respective mass absorption coefficients in
cm$^2$/g, for an X-ray with energy $E_X$,  $E_{K}^{Ar}$ is the
shell K ionization potential for argon, $P_{esc}$ the escape
probability of the fluorescence X-ray produced by radiative
deexcitation of the K-shell, and $w$ the average energy for
producing one ionization pair in the respective gas. Finally the
thickness $x_{gas}$ of the respective gas in g/cm$^2$ is given by
the well known formula
\begin{eqnarray}
x_{gas} & = & \frac{m_{gas}\, p_{gas}}{R\, T}\,\langle t\rangle
\end{eqnarray}
being $\langle t\rangle$ the average tracklength of the X-ray in
the gas, $m_{gas}$ the molecular weight of the gas, $p_{gas}$ its
partial pressure, $T$ the absolute temperature and $R$ the gas
constant. We observe that a small variation of the pressure and/or
temperature of the gas is amplified by the exponential dependance
in Eq.\ (\ref{eq:current}). In order to correct for small changes
of the pressure and temperature of the gas mixture during the run,
we have monitored the pressure of the gas with an accuracy of
$\pm0.5$ mb and the temperature of the test tube with an accuracy
of $\pm0.1°$ C.

\begin{figure*}[htb]
\centering
\includegraphics[width=13cm, height=8cm]{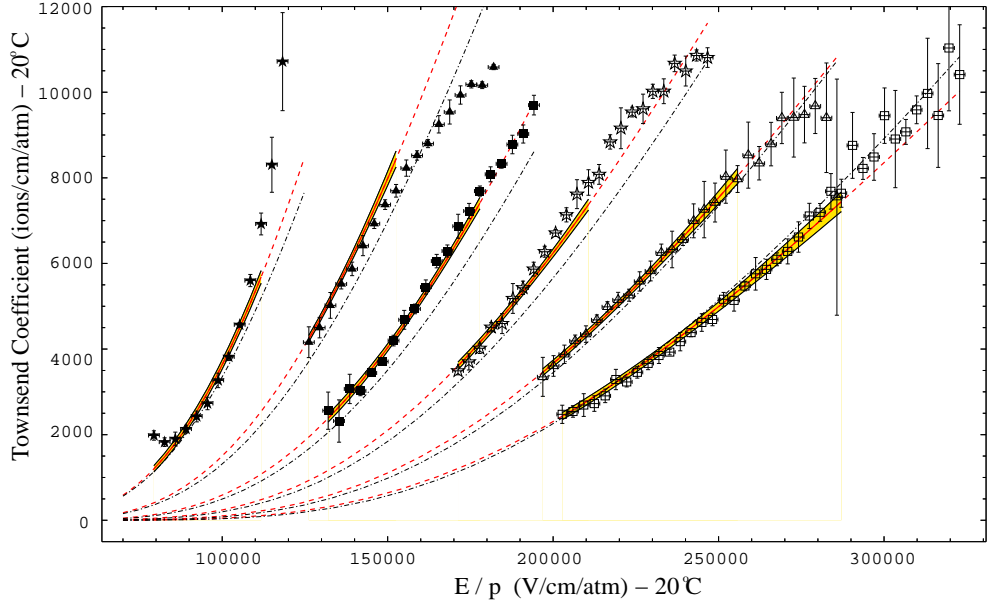}
\caption{Townsend coefficient vs.\@ reduced electric field. The
six curves are obtained for Ar-CO$_2$ mixtures spanning from
100-0\% to 0-100\% relative percentages. The thin dot-dashed lines
are the predictions of Eq.\ (\ref{eq:disc3}) with the parameters
$A$ and $B$ have been fitted separately for the pure gases. The
heavy dashed lines are obtained from a two-dimensional fit of the
entire data set with the same formula (see text).}
\label{fig:townfit}
\end{figure*}

\section{Experimental results}\label{sect:expres}

We report here the results obtained in five different experimental
runs, repeated in order to check the reproducibility of the
measurements. Each run has been performed in completely automatic
mode. We have adopted a rather complex strategy after some
preliminary measurements, that have given to us a feeling for the
various dynamical time scales involved in the system. First we
have observed that the flow of the gas stabilizes at the assigned
partial flows only after a period of at least half a day. During
this period we found that a constant applied voltage of 1000 V
helps the conditioning of the test tube to the new mixture. After
this time the anode voltage is changed by the program to the
minimum voltage and raised in steps of 25 V, with slow ramp of 2
V/s. When the high voltage has reached the programmed value, the
program allows for additional 180 s waiting time, before starting
the current measurements at this voltage. The measurements are
continued until the cathode current is below 2 nA, which is in our
experience the level of self sustained discharge. We have checked
that in this way no sequentiality in the measurements is observed.

The anode current is measured integrating over a time window
depending on the picoammeter fullscale. For the majority of the
performed measurements the integration time of the picoammeter was
$\sim$600 $\mu$s. As a consequence the single current measurement
is affected by large poissonian fluctuations. The statistical
error has been reduced to be negligible, compared with the
calibration and linearity error of the picoammeter, averaging over
300 single measurements of the current. Therefore we have assigned
to each current measurement an error corresponding to 0.4\% of the
measured value, plus a zero point uncertainty of $\pm
0.5\;\mathrm{pA}$. Actually the LSD of the read out of the
instrument corresponds to $\pm 0.05\;\mathrm{pA}$, but we have
observed that the zero point current reproducibility is not at
this level, likely due to some unavoidable parasitic current.

From the values of the current we have obtained the gas gain
dividing by the calculated primary current. Then we have corrected
the calculated gain for the effect of small variations in the
primary ionization current according to Eq.\ (\ref{eq:current})
using the pressure and temperature data, monitored for each
measurement. From these values the estimated $\alpha$ coefficient
has been obtained, by evaluating the derivative by discrete
incremental ratios. In order to evaluate the robustness of the
method against possible numerical instabilities, we have also
checked that the discrete values obtained in this way are
compatible with the smooth curve obtained from the differentiation
of a polynomial fit to the data. Finally we have computed the
effective anode reduced field and the value of the $\alpha$
coefficient using the actual pressure measurement, scaled to a
reference temperature of $20^\circ$ C.

The entire cycle of measurements and the data reduction has been
repeated six times, spanning from Ar-CO$_2$ 100-0$\%$ to 0-100\%
in steps of 20\%. We have observed that each set of curves is
fully compatible inside the experimental error. In Fig.\
\ref{fig:townfit}, which constitutes the central result of this
work, we have reported the average of the values obtained in the
five runs.

\begin{table}[ht!]
\begin{center}
\begin{tabular}{|c|c||c|}
\hline& Pure gases & Combined fit  \\ \hline
A$_1$  & $(1.38 \pm 0.07) \times 10^5  $& $(1.07 \pm 0.08) \times 10^5$ \\
B$_1$ &  $(6.05 \pm 0.08) \times 10^4$  & $(5.6 \pm 0.1) \times
10^4 $
\\\hline
A$_2$ & $(2.2 \pm 0.9) \times 10^5$ & $(2.6 \pm 0.2) \times 10^5 $  \\
B$_2$  & $(2.6 \pm 0.3) \times 10^4 $ & $(2.65 \pm 0.05) \times
10^4 $\\ \hline
\end{tabular}
{\footnotesize {} }
\end{center}
\caption{In the first column the parameters to be inserted in
Eq.\@ \ref{eq:disc1} obtained from two separate fits of the data
for pure gases only. The second column shows the results obtained
from the fit of the model of Eq.\@ (\ref {eq:disc3}) over the
entire data set. Subscript 1 refers to $CO_{2}$, and 2 to $ Ar$.
Errors are only the statistical error of the fit.}
\label{tab:additive}
\end{table}
\section{Discussion}

\label{sect:discus}

 Several parameterizations of the Townsend
coefficient have been proposed in the literature (for a review see
e.g. Ref.\cite{Aoyama:1985} and references therein).
\begin{figure}[!ht]
\centering
\includegraphics[width=8cm]{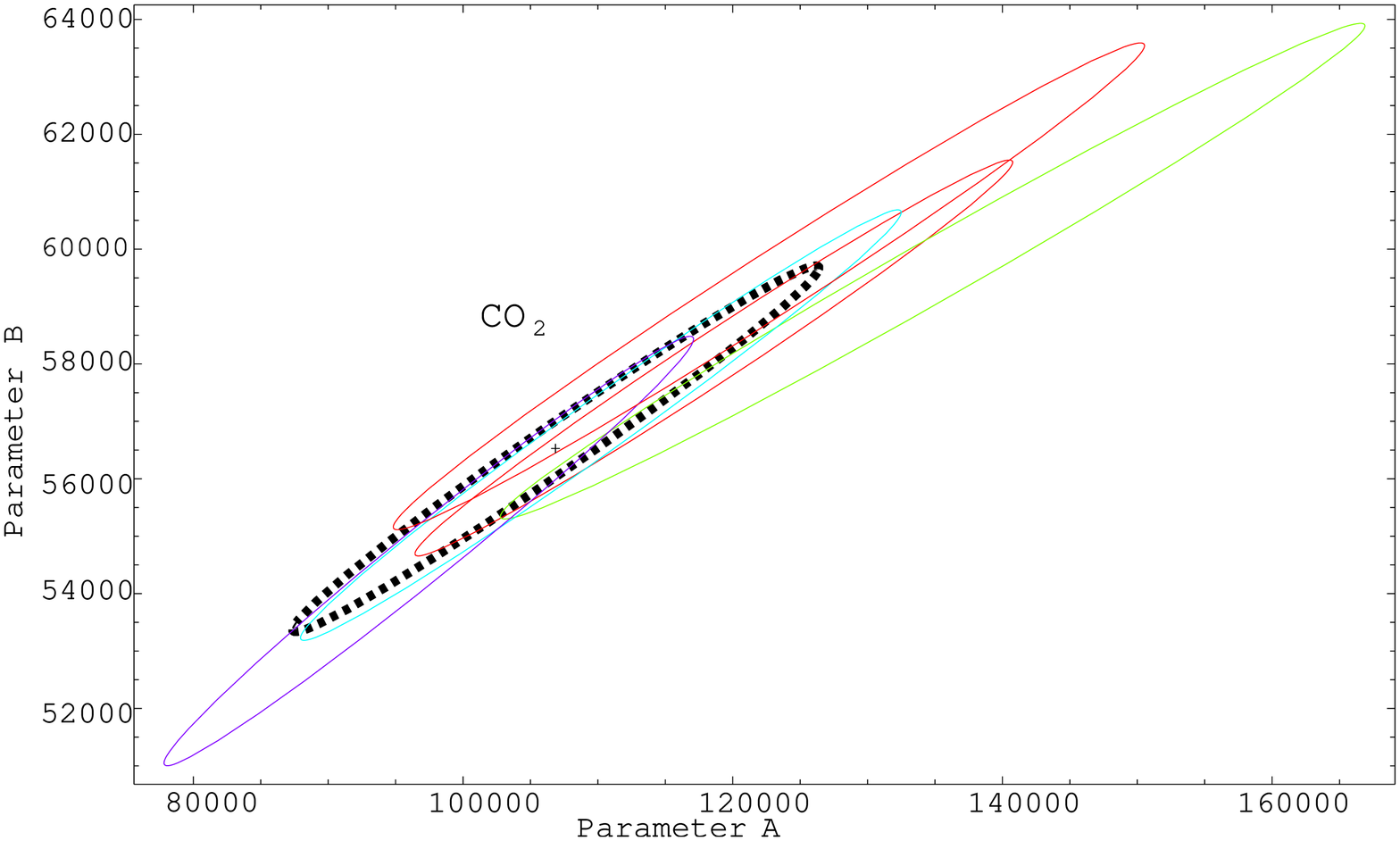}
\caption{Joint 95\% confidence regions for the best fitted
parameters relative to $CO_{2}$. The thicker line refers to the
results for the averaged data.} \label{fig:towncomp1}
\end{figure}

\begin{figure}[!ht]
\centering
\includegraphics[width=8cm]{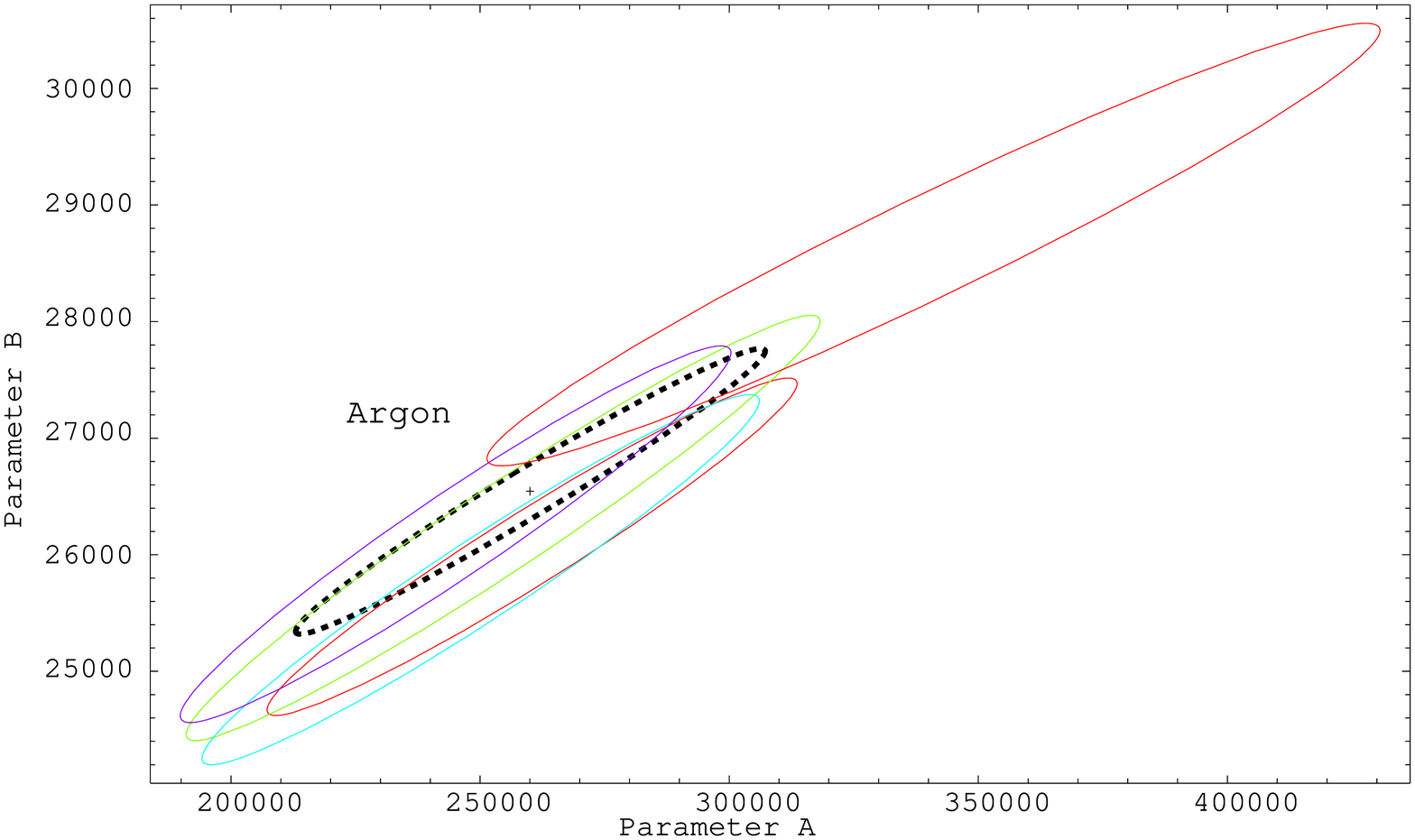}
\caption{Joint 95\% confidence regions for the best fitted
parameters relative to Argon. The thicker line refers to the
results for the averaged data.} \label{fig:towncomp2}
\end{figure}

However a simple thermal distribution of the electron energies,
leads to a function of the type\cite{Williams:1962}
\begin{equation}
\alpha (S)=A\,e^{-\epsilon_{ion}B/S}  \label{eq:disc1}
\end{equation}
which has the appeal of a straightforward physical interpretation.
In fact if we assume that the accelerated electrons have
maxwellian velocity distribution with temperature $kT_{e}\approx
e\,E\,\lambda _{coll}$, the ionization rate from ground state of
the gas is \cite{McWhirter:1965} $\propto \exp
[-\epsilon _{ion}/kT_{e}]$ giving $B\approx \langle 1/e\,\lambda _{coll}\rangle$, while $%
A=\langle1/\lambda _{ion}\rangle$ where $\lambda _{coll}$ is the
 m.f.p. of the electron for elastic and inelastic scatterings,
 while $\lambda _{ion}$ is the m.f.p. for ionizing scatterings only.
 In this context the $\langle\rangle$ brackets indicate an average
 of the energy dependent m.f.p. over the relevant energy range of the electrons.

It is remarkable that our data for pure Argon and pure CO$_2$,
reported in Fig.\ \ref{fig:townfit}, can be well fitted with Eq.\@
(\ref{eq:disc1}) in the lower part of the reduced field range, up
to a value of $\alpha\approx 8,000\;\mathrm{ions/cm/atm}$. The
fitted value are reported in the second column of Tab.\
\ref{tab:additive}.

Limiting our attention to mixtures of two gases, we propose here a
simple additive model in which the ionization densities produced
in the mixed gas is the sum of the densities in each gas.
Therefore, starting from Eq. (\ref{eq:disc1}) for the pure gases,
we can easily write for a mixture
\begin{equation}
\alpha =p_{1}\,A_{1}\,e^{-\epsilon
_{1}B^*/S}+p_{2}\,A_{2}\,e^{-\epsilon _{2}B^*/S} \label{eq:disc3}
\end{equation}
where $B^*=(p_{1}B_{1}+p_{2}B_{2})$, $p_{1}$ and $p_{2}$ are the
relative partial pressures, $\epsilon_{1}$ and $\epsilon _{2}$ the
first ionization potential of the two gases.

In Fig.\ \ref{fig:townfit} we have reported as thin dot-dashed
lines, the curves obtained from Eq.\ (\ref{eq:disc3}), in which we
have inserted the coefficients $A$ and $B$ fitted on the data from
pure gases only. It can be seen that the estimates of the Townsend
coefficient for the various mixtures obtained in this way are not
far from the measured values.

More interesting is the result that we have obtained from a
two-dimensional fit of the entire data set with the model of Eq.\@
(\ref{eq:disc3}), leaving as free parameters the coefficients $A$
and $B$ of the pure gases, and considering the reduced field and
the partial pressures independently measured coordinates of the
data points. In Fig.\@ \ref{fig:townfit} we show the result of the
fit in this case as heavy dashed lines. It is clear that in this
way we obtain a more reliable fit of the coefficients of the pure
gases, because we use all the available information on the $A$ and
$B$ coefficients for the pure gases at the same time. The fitted
parameters are reported in the third column of Tab.\
\ref{tab:additive}. We have also performed a separate fit to the
curves obtained from of each run. As can be seen from Figs.\@
\ref{fig:towncomp1} and \ref{fig:towncomp2}, where the 95\%
confidence regions are reported, we also find that all of the fits
of the single runs are compatible within $2 \sigma$.

We conclude that our work shows that our proposed method of
estimating the Townsend effective coefficient from the slope of
the $\ln M$ vs. $\ln V$ curves is experimentally robust. It also
shows that the simple functional form of Eq.\ (\ref{eq:disc1}) is
adequate for predicting the evolution of the Townsend coefficient
at high fields, if the detector is far from the regime of
self-sustained regenerative discharge. In spite of its simplicity
the additive model for the Townsend coefficient, in the form of
Eq.\ (\ref{eq:disc3}) can be reliably used to predict the behavior
of Ar-CO$_2$ mixture. It will be interesting to extend in the
future this type of investigation to more complex mixtures,
including gases with strong electron attachment and photon
regeneration such as for example the CF$_4$.

\end{document}